\documentclass[aps,prd,twocolumn,superscriptaddress]{revtex4-1}

\usepackage{graphicx}
\usepackage{color}

\newcommand{\Linit}{L_{\rm init}}
\newcommand{\Erad}{E_{\rm rad}}
\newcommand{\Lrad}{L_{\rm rad}}
\newcommand{\msun}{M$_{\rm \odot}$}
\newcommand{\dxl}{{\rm dx}_{\rm L}}
\newcommand{\dxm}{{\rm dx}_{\rm M}}
\newcommand{\dxh}{{\rm dx}_{\rm H}}
\newcommand{\bcrit}{b_{\rm crit}}

\newcommand{\mnras}{Mon. Not. R. Astron. Soc.}

\newcommand{\apjl}{Astrophys. J. Lett.}

\begin{document}

\title{Gravitational Radiation Driven Capture in Unequal Mass Black Hole Encounters}

\author{Yeong-Bok Bae}
\email{baeyb@astro.snu.ac.kr}
\affiliation{Astronomy Program Department of Physics and Astronomy, Seoul National University, 1 Gwanak-ro, Gwanak-gu, Seoul  08826, Korea}
\affiliation{Korea Astronomy and Space Science Institute, 776 Daedeokdae-ro, Yuseong-gu, Daejeon 34055, Korea}

\author{Hyung Mok Lee}
\email{hmlee@snu.ac.kr}
\affiliation{Astronomy Program Department of Physics and Astronomy, Seoul National University, 1 Gwanak-ro, Gwanak-gu, Seoul  08826, Korea}
\affiliation{Center for Theoretical Physics, Seoul National University, 1 Gwanak-ro, Gwanak-gu, Seoul 08826, Korea}

\author{Gungwon Kang}
\email{gwkang@kisti.re.kr}
\affiliation{Supercomputing Center at KISTI, 245 Daehak-ro, Yuseong-gu, Daejeon 34141, Korea} 

\author{Jakob Hansen}
\email{jakob@jakobonline.dk}
\affiliation{Supercomputing Center at KISTI, 245 Daehak-ro, Yuseong-gu, Daejeon 34141, Korea}

\date{\today}

\begin{abstract}

The gravitational radiation driven capture (GR capture) between unequal mass black holes without spins has been investigated with numerical relativistic simulations. We adopt the parabolic approximation which assumes that the gravitational wave radiation from a weakly hyperbolic orbit is the same as that from the parabolic orbit having the same pericenter distance. Using the radiated energies from the parabolic orbit simulations, we have obtained the critical impact parameter ($\bcrit$) for the GR capture for weakly hyperbolic orbit as a function of initial energy. The most energetic encounters occur around the boundary between the direct merging and the fly-by orbits, and can emit several percent of initial total ADM energy at the peak. When the total mass is fixed, energy and angular momentum radiated in the case of unequal mass black holes are smaller than those of equal mass black holes having the same initial orbital angular momentum for the fly-by orbits. We have compared our results with two different Post-Newtonian (PN) approximations, the exact parabolic orbit (EPO) and PN corrected orbit (PNCO). We find that the agreement between the EPO and the numerical relativity breaks down for very close encounters ($\it{e.g.}$, $\bcrit \lesssim 100$ M), and it becomes worse for higher mass ratios. For instance, the critical impact parameters can differ by more than $50\%$ from those obtained in EPO if the relative velocity at infinity $v_{\infty}$ is larger than 0.1 for the mass ratio of $m_{1}/m_{2}=16$. The PNCO gives more consistent results than EPO, but it also underestimates the critical impact parameter for the GR capture at $\bcrit \lesssim40$ M. 

\end{abstract}



\maketitle

\section{\label{intro}Introduction}

Binary black hole (BBH) merger has been regarded as one of the most promising sources of gravitational waves (GWs) \citep{banerjee10,downing10,downing11,morscher13,tanikawa13,bae14}. In fact three such events were eventually observed by the advanced LIGO \citep{abbott1,abbott2,abbott17} recently. One of the formation processes of the compact BBH is a gravitational radiation driven capture (GR capture, hereafter) which occurs from close encounters between two black holes (BHs) in very dense stellar systems such as globular clusters and galactic nuclei \citep{oleary09,hong15}. Capture takes place when the amount of energy radiated by gravitational waves during the encounter becomes larger than the initial orbital energy.

Most studies on GR capture are based on Post-Newtonian (PN) order of 2.5 (2.5 PN) by \citet{peters64} and \citet{hansen72}. \citet{peters64} calculated the amount of gravitational radiation from two point masses assuming bound Keplerian orbits, and \citet{hansen72} extended this work to hyperbolic Keplerian orbits. \citet{quinlan87,quinlan89,mouri02} obtained the cross section of GR capture based on PN results. The GR capture in dense star cluster \citep{lee93,oleary09,hong15}, and recent studies about the primordial BBH merger \citep{bird16,mandic16,clesse16,clesse17} employed cross section from PN. 

PN calculations of GR capture can be applied for most cases of BH encounters whose orbital energy is much smaller than the rest-mass energy, i.e., non-relativistic cases. The amount of energy that has to be dissipated for the capture is determined by the velocity dispersion of the cluster, because typical orbital energy of two unbound stars is approximately $(3/2)\mu\sigma^2$ where $\mu$ is the reduced mass and $\sigma$ is the one-dimensional velocity dispersion of the cluster. The velocity dispersion for a typical star cluster such as globular cluster or open cluster is less than a few km s$^{-1}$, and a few hundred km s$^{-1}$ for the central parts of the galaxies. Therefore PN approximation is generally sufficient for GR captures in those clusters.

However, if the velocity dispersion of the cluster becomes very large, the full relativistic treatments are required because the PN approximation becomes inaccurate. Most of the galaxies are known to harbor supermassive BHs whose masses are proportional to the masses of host galaxies (e.g., \citep{ferrarese00,gillessen09,genzel10,kormendy13}). Stars around the BH inside the radius of influence follow cuspy density distribution of $r^{-7/4}$ for the case of single mass \citep{bahcall76}. The velocity dispersion also follows a power-law of $r^{-1/2}$, implying the possibility of very large value at the very small distances from the central supermassive BH. If many stellar mass BHs are concentrated around the supermassive BH \citep{morris93,miralda00,freitag06,hopman06} and they are composed of various masses, the density profile of more massive components is expected to be steeper than the case of single mass (e.g., \citep{alexander09}). Therefore, the encounter velocities between stellar mass BHs in the galactic nuclei could be very large. The formation process of such extreme binaries requires full general relativistic numerical simulations which became available after the breakthrough of long-term stable evolution \citep{pretorius05,campanelli06,baker06}.

One of the interesting consequences of GR captures is the possibilty of forming very eccentric binaries. Note that most BBHs are expected to be circularized before the merging \citep{peters64,oleary06} due to the loss of the orbital energies during the inspiral phase. Therefore the analyses of GW data and the parameter estimations of GW sources have focused on the binaries with circular orbits. However, the typical deviation of eccentricity from 1 of the captured binaries can be of order of $1-e \approx 6.5 \times 10^{-3} {\eta}^{2/7} {(\sigma/1000~\rm{km~s^{-1}})}^{10/7}$ where $\eta\equiv m_{1}m_{2}/(m_{1}+m_{2})^2$ is the symmetric mass ratio \citep{hong15}. On the other hand, typical pericenter of the capured binaries is $r_p \approx 5.7\times 10^{2} (\sigma/1000~\rm{km~s^{-1}})^{-4/7}~\rm{km}$ for the 10 \msun~BH pairs \citep{hong15}. Thus, if the velocity dispersion is $\sigma=1000~\rm{km~s^{-1}}$, the pericenter of the captured binary is only about 20 Schwarzschild radii. Consequently the captured binaries in dense stellar systems with high velocity dispersion can radiate GWs and merge while possessing significant eccentricity. Therefore, the waveform modeling and the parameter estimation of eccentric orbit are considered only recently (e.g.,\citep{sun15,huerta17}).

The purpose of the present study is to extend the numerical simulations of the close encounters between two unequal mass BHs. The previous study by Hansen $\it{et \, al.}$~\cite{HKKDL} made very detailed numerical simulations of GR capture between equal mass BHs. Since the observed masses of the stellar mass BHs span from about 5 to 30 \msun~as evident from the X-ray binaries and GW150914, encounters between unequal mass BHs would be more common than equal mass ones.

In this study, the geometrized unit system is adopted to express the physical quantities. The speed of light in vacuum $c$ and the gravitational constant $G$ are set to be unity in the geometrized system, and the length, time, mass and energy are expressed with the units of mass M.

The outline of this paper is as follows. In section \ref{setup}, the details of our relativistic simulations---the assumptions,   the software we have adopted and the configurations of initial conditions etc.---are described. In section \ref{convg}, the convergences of our simulations are tested. The main results of our simulations about GR capture between unequal mass BHs are presented in section \ref{results}. In section \ref{validity}, the validity of parabolic approximation we employ is examined, and in section \ref{discuss} the application of relativistic GR capture is discussed.


\section{\label{setup}Numerical Setup}

In order to simulate the GR capture process, we need to follow hyperbolic orbits numerically.  There are two conservative quantities for hyperbolic orbits in non-relativistic limit: orbital energy and angular momentum. Since we do not know these quantities for the orbits that lead to the capture, large number of full simulations are required. However, it would be difficult to do so with limited computational resources. 

Instead of carrying out a large number of numerical simulations for initially hyperbolic orbits, we adopt the parabolic approximation that can reduce the parameter space we should search since orbital energy is zero for parabolic orbits. Parabolic approximation is based on the assumption that the weakly hyperbolic orbit emits the same amount of energy with the parabolic orbit of the same pericenter distance, because both orbits have almost same paths around the pericenter where most GWs are radiated. Under the parabolic approximation, we first calculate the gravitational radiation energy of parabolic orbits as a function of the pericenter distances. Any hyperbolic encounters with the orbital energy smaller than the radiated energy for a given pericenter distance are considered to lead the capture.



In Newtonian limit, the parabolic orbit can be realized by setting total energy as the sum of the rest masses because its orbital energy is zero. Similarly, the initial parabolic orbit in this study is obtained by constraining total ADM mass as the sum of ADM masses of each puncture \citep{ansorg04}. The ADM mass \citep{adm62} is defined in the asymptotically flat space-time as the total mass-energy measured within a spatial infinite surface at any instant of time, thus it corresponds to total energy in Newtonian limit. In our simulations the corresponding value of total ADM mass is 1.

We use Einstein Toolkit \citep{loffler12,toolkit} which is open to the public for fully general relativistic simulations, and adopt the standard thorns and settings (1+log lapse, Gamma-driver shift etc.). The thorn `McLachlan' \citep{brown09} is used for the vacuum spacetime solver and the apparent horizons of BHs are tracked by the `AHFinderDirect' thorn \citep{thornburg04}. 

The initial data are constructed as follows. 
\begin{enumerate}
\item Two BHs are located on x-axis with a certain initial separation ($d$) such that they have equal and opposite linear momenta.
\item We choose initial angular momentum ($\Linit$), i.e., y-directional linear momentum ($p_{y}$) of each BH ($\Linit=d p_{y}$).
\item For a given $\Linit$, we choose an arbitrary x-directional linear momentum ($p_{x}$) and let TwoPuncture thorn \citep{ansorg04} find bare masses such that two BHs have the desired ADM masses.
\item We measure total ADM mass. If it is larger than 1, we decrease $p_{x}$. Conversely we increase $p_{x}$ if total ADM mass is smaller than 1. 
\item We repeat step 4 until the total ADM mass is close enough to 1 with the maximum tolerance $10^{-10}$.
\end{enumerate}
Here we used sufficiently large initial separation ($d=60$ M) between two BHs in order to reduce the junk radiation and distinguish it from true GW signals. The `mass ratio' in this study is defined as the ratio between the ADM masses of the BHs (Eq. (83) in \citep{ansorg04}). For example, we set the ADM mass of each BH as `0.8' and `0.2' respectively for the mass ratio 4. However, the total ADM mass may not be unity because the total ADM mass is not just the sum of two ADM masses. Thus, we need some iterations (step 5) to find the parabolic orbit. Note that the puncture mass ratio between two BHs can be slightly different from the ADM mass ratio.

In order to get the accurate radiated energy and angular momentum, GWs should be extracted at sufficiently large distances where the BH system can be recognized as a point mass. For the parabolic orbit with larger pericenter distance, the extraction radii have to be correspondingly larger because the effective size of the BH system is larger. It requires larger simulation domain. Additionally, the unequal mass system does not have 180 degree rotational symmetry unlike the equal mass system. This means that the simulation domain has to be twice of the equal mass system. Spatial resolution should also be more stringent for unequal mass case than equal mass case since the minimum spatial grid has to be detemined by the lower mass BH. Consequently, much more computer resources are required for the simulations of the unequal mass BHs than those of the equal mass BHs.

To overcome such a resource problem, we adopt the multiblock infrastructure, Llama code \citep{pollney11}. In a Cartesian coordinate, we can use a mesh refinement based on `Carpet' \citep{schnetter04} within Cactus. Unfortunately the expansion of the simulation domain is very expensive because the number of grid points is proportional to the cube of the length scale. However, if we can use the spherical-like coordinates, the required number of grid points is just proportional to the length scale, provided that the number of angular grid points is fixed. The Llama code \citep{pollney11} enables us to use both coordinates in the simulation domain. The Cartesian coordinates are adopted for the central region where the mesh refinement is required around the BHs, while the spherical-like coordinates are used for the outer region where the GWs are extracted. Information on different patches is shared by the interpolation in the overlapping grids. Consequently, we can expand the simulation domain much farther with fewer computational resources. 

In this study, we adopt 7 patches system which is composed of central Cartesian coordinates and 6 spherical-like patches attached on faces of the Cartesian cube (See Fig. 1 in \citep{pollney11}). In the Cartesian coordinates, the mesh refinement is applied for each BH with 7--10 refinement levels according to the required resolutions. Unlike the quasi-circular merging simulations, two BHs can be separated far away after the encounter, thus the central patch where we can use the mesh refinement around the BHs has to be sufficiently large to encompass two BHs. Inner radii of spherical-like coordinates are set to 80 M--120 M depending on the cases. Outer boundaries of spherical-like coordinates are set to $\gtrsim1200$ M to avoid the errors caused by the wave reflection from the outer boundary.

In the previous study of equal-mass BHs \citep{HKKDL}, the finest grid size was $1.1/64$ M. However, higher resolutions are required for the lower mass BH in the unequal mass BH simulations. In this study, the finest grid sizes have been determined by considering the puncture mass and the apparent horizon size in an inversely proportional manner, and have been set to $1.25/128$ M, $1.5/256$ M, $1.8/512$ M and $1.2/512$ M for the mass ratios of 2, 4, 8 and 16 respectively. We have used relatively low resolution for $m_{1}/m_{2}=16$ due to the limited computer resources, but it is also in the convergent regime as shown in section \ref{convg}.

Weyl scalar $\Psi_{4}$ in Newman-Penrose formalism is extracted using WeylScal4 thorn, and decomposed into the spin-weighted spherical harmonics with Multipole thorn \citep{baker02}. Ideally the GWs should be extracted at infinity, and Cauchy-Characteristic extraction code \citep{babiuc11,bishop99} in Einstein Toolkit can be one of the options for that. In this study, however, we followed standard way of calculating the radiated energy and angular momentum by extrapolating the integration of GWs from several extraction radii. Farther extraction radii are used for large initial angular momentum cases because they have larger pericenter distance. Depending on the pericenter distances, the extraction radii of 120 M--500 M are used in the simulations.


\section{Convergence Test\label{convg}}

\begin{figure*}
\includegraphics[width=1\textwidth]{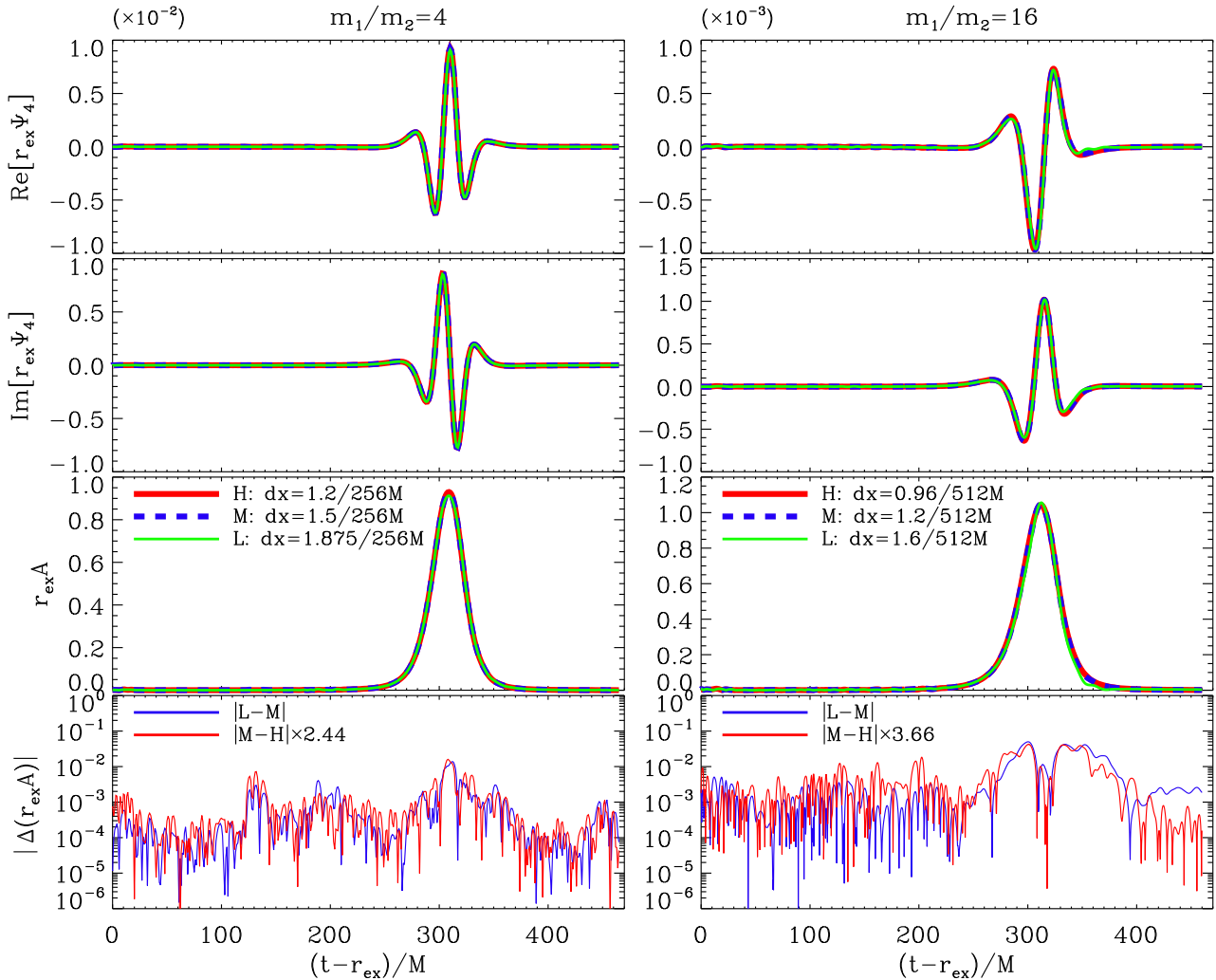}
\caption{Comparison of $\Psi_{4}(l=2,m=2)$ and amplitudes for three different resolutions with time. The retarded time ($t-r_{\rm ex}$) is used because it takes time for GWs to arrive at the extraction radii. Left panels show the case of $m_{1}/m_{2}=4$ and the initial angular momentum is $L_{\rm init}=0.64$, and Right panels for $m_{1}/m_{2}=16$ and $L_{\rm init}=0.24$. The lowest panels are the absolute values of the differences of $\Psi_{4}(l=2,m=2)$ between the different resolutions. In both cases `$|$M-H$|$' is scaled for 4th-order convergence and it overlaps the `$|$L-M$|$' in the time range where the GWs are radiated. \label{converge}}
\end{figure*}

In order to check whether we are using the appropriate resolutions, the convergence should be tested with the different grid sizes. Since we have various models with different mass ratios and initial angular momenta, it is not possible to check the convergences for all the models. We have selected two representative cases of close fly-by orbits with the mass ratio $m_{1}/m_{2}=4$ (initial angular momentum $L_{\rm init}=0.64$) and 16 ($L_{\rm init}=0.24$). Three different resolutions dx$=1.875/256$ M, $1.5/256$ M and $1.2/256$ M are adopted for the mass ratio $m_{1}/m_{2}=4$, and dx$=1.6/512$ M, $1.2/512$ M and $0.96/512$ M for $m_{1}/m_{2}=16$ where `dx' is the finest grid size in the mesh refinement encompassing the BHs. Hereafter, the letters `L', `M' and `H' represent the low, medium and high resolutions, respectively. The box sizes of mesh refinement levels are kept identical among the different resolutions.

Figure~\ref{converge} shows the real and imaginary parts of the Weyl scalar $\Psi_{4}$ in ($l,m$)=(2,2) mode, and the amplitudes of them. For the comparison, the $\Psi_{4}$ and amplitudes are multiplied by the extraction radius $r_{\rm ex}$, and measured at the retarded time. Since they show almost the same behaviours, we have subtracted one from the other to see the differences in detail. The lowest panels show the absolute values of the differences of $\Psi_{4}$ between `L' and `M' ($|$L-M$|$), and `M' and `H' ($|$M-H$|$). The `$|$M-H$|$' is multiplied by 2.44 (for $m_{1}/m_{2}=4$) and 3.66 (for $m_{1}/m_{2}=16$) for scaling with 4th-order convergence and it overlaps the `$|$L-M$|$' in the time range where the GWs are radiated. Here, the factors 2.44 and 3.66 are obtained from the grid size ratios among different resolutions, i.e., 
\begin{equation}
\frac{(\dxl/\dxm)^4 (\dxm/\dxh)^4 - (\dxm/\dxh)^4} {(\dxm/\dxh)^4-1}
\end{equation}
where $\dxl$, $\dxm$ and $\dxh$ is the finest grid size of `L', `M' and `H'. Therefore, the 4th-order convergence is confirmed in that time range. 

Ideally, our simulations should have had 8th-order convergence because we adopted 8th-order finite difference method. However, we get the lower order of convergences due to the practical reasons---lower order operations on boundaries (mesh refinement, multi-patches, etc.). Our simulations have about 4th-order convergences which are consistent with \citet{pollney11}.

The medium resolutions in the convergence test are adopted in the following simulations, because it is considered to be sufficient for this study. The difference of the radiated energies between the medium and high resolutions is about 0.5\% for the mass ratio $m_{1}/m_{2}=4$, and about 1.5\% for $m_{1}/m_{2}=16$. 

Note that the convergence tests are just for the representative models. However, the resolutions for the other models are also adjusted in similar manner based on the puncture mass and the size of the apparent horizon as stated in section \ref{setup}.


\section{\label{results}Results}

We have performed the parabolic orbit simulations for the different mass ratios up to $m_{1}/m_{2}=16$. Since the ADM mass of the parabolic orbit is fixed, we only need to specify the initial angular momentum ($\Linit$) to describe the orbit for a given mass ratio. Here, the initial angular momentum is calculated by $\Linit= x_{1}p_{y,1}+x_{2}p_{y,2}$ where $x_{1}$, $x_{2}$ are the positions of two BHs on the x-axis and $p_{y,1}$, $p_{y,2}$ are the linear momenta in y-directions. 

\begin{figure}
\includegraphics[width=0.95\columnwidth]{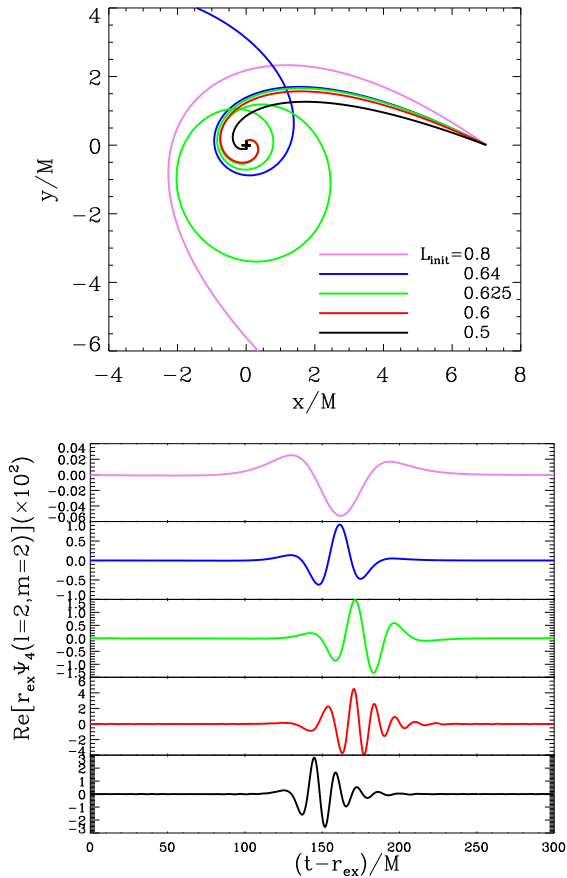}
\caption[Orbits and waves for different initial angular momenta when the mass ratio $m_{1}/m_{2}=4$]{Upper panel shows the orbit of heavier BHs with different initial angular momenta when the mass ratio $m_{1}/m_{2}=4$. We have displayed the orbit starting from when the separation between BHs is 35 M (arbitrarily chosen). Lower panel is the real part of $\Psi_{4}$ multiplied by the extraction radii ($r_{\rm ex}$). Same colors represent the same $\Linit$. \label{orbit_wave}}
\end{figure}

Several examples are shown in Fig.~\ref{orbit_wave} with the mass ratio of $m_{1}/m_{2}=4$. Upper panel shows the orbits of heavier BHs. Although the initial separation of BHs is 60 M in all simulations, we have displayed the orbit starting from when the separation is 35 M (arbitrarily chosen) and rotated each plot such that the starting point is (7 M, 0). The orbits of lower mass BHs, in the opposite side of the higher mass with four times larger distances, are not shown for simplicity. Lower panel shows $\Psi_{4}$ from those orbits.

For large $\Linit$, two BHs show fly-by orbit without merging. They are in bound state, but their apocenter is too far to be simulated. As the $\Linit$ decreases, BHs end up with more tightly bound orbits (e.g., $\Linit=0.625$). BHs merge nearly directly for two other cases of smaller $\Linit$ (0.6 and 0.5). Note that the scales of the vertical axes in the lower panel are different from each other. As the $\Linit$ decreases, the amplitude of $\Psi_{4}$ becomes larger, and it has the largest value on the boundary between the fly-by and direct merging orbits. If we decrease $\Linit$ below the boundary, the amplitude becomes smaller again.

In this study, we want to obtain the marginal energy of the GR capture through the parabolic simulations. Parabolic orbit is used just to obtain the marginally capturing hyperbolic orbit. The information we need is the radiated energy from just one-passage in the parabolic orbit. Therefore, the direct merging orbit cannot be used for calculating the marginal energy of GR capture process.


\subsection{Radiated Energy \& Angular Momentum \label{radEL}}

\begin{figure}
\includegraphics[width=0.95\columnwidth]{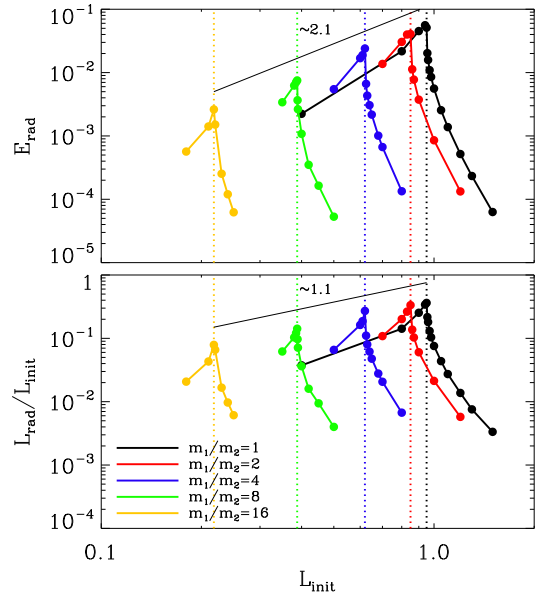}
\caption[Radiated energy and angular momentum with different mass ratios]{The radiated energy $\Erad$ and angular momentum $\Lrad$ with different mass ratios. $\Lrad$ is divided by the initial angular momentum $\Linit$ to represent the radiation fraction. $\Erad$ is the radiation fraction itself, because the initial ADM energy is 1. The dotted lines are the boundaries between the direct merging and the fly-by orbit. Thin black solid lines represent the linear fitting of the peak points with their slopes in logarithmic scale. \label{nr_radEL}}
\end{figure}

We have measured the radiated energies ($\Erad$) and angular momenta ($\Lrad$) of encounters on parabolic orbits with different mass ratios by integrating $\Psi_{4}$ up to $l=12$ modes in multipole expansions. However, high l-modes are not in fact critical to this calculations. We have examined the case of the mass ratio $m_{1}/m_{2}=16$ with $L_{\rm init}=0.24$ up to $l=16$ modes, and found that $l=(9-16)$ modes contribute only about 0.1\% to the total radiated energy. Discarding modes beyond l=12 seems to lead an error no larger than that, and is therefore acceptable. We have also confirmed that the error from each mode is smaller than the energy itself even in the high l-modes.

Figure~\ref{nr_radEL} shows $\Erad$ and $\Lrad$ as a function of the initial angular momentum ($\Linit$) for five different mass ratios. There are peak points both in $\Erad$ and $\Lrad$ for each mass ratio. In the left hand side of the peak, including the peak point itself, two BHs merge directly. The $\Erad$ and $\Lrad$ in this part are integrated values up to the merger. On the other hand, the right hand side of the peak is from the fly-by orbits, where we can calculate $\Erad$ and $\Lrad$ for the entire one-passage. Regardless of the mass ratio, the orbit on the boundary between the direct merging and the fly-by gives the largest amount of GW radiations. As the initial angular momentum deviates further from this boundary, the GW radiation drops quickly, especially for the fly-by orbits. 

As the mass ratio becomes large, the boundary between the direct merging and the fly-by orbits shifts to the lower initial angular momentum (Fig.~\ref{nr_radEL}). These $\Linit$ at the peak are roughly proportional to the reduced mass $\mu=m_{1}m_{2}/(m_{1}+m_{2})$, implying that the specific initial angular momenta at the peaks are almost the same. In addition, the radiated energies and angular momenta at the peak follow the power law with $\Erad$, $\Lrad\propto\Linit^{2.1}$. However, we should be careful in extrapolating these results to the higher mass ratio, because the peak points cannot be defined precisely. Near the peak points, BHs can have whirling orbits around each other. In such cases it is difficult to define the one-passage because two BHs approach again each other before having sufficient distance. 

\begin{figure}
\includegraphics[width=0.95\columnwidth]{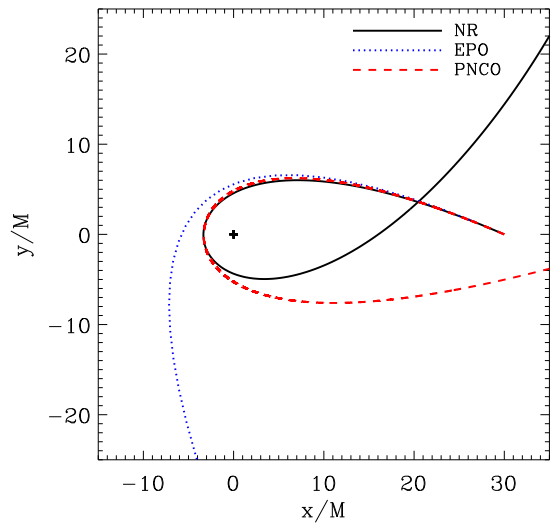}
\caption[The comparison of the EPO, PNCO and NR orbit]{The comparison of the exact parabolic orbit (EPO), PN corrected orbit (PNCO) and the orbit from numerical relativity (NR). The mass ratio is 1 and the initial angular momentum is $\Linit=1.1$. Here, we present the orbit of one BH. The other is opposite side of the origin.\label{pn_orbit}}
\end{figure}

The results of the numerical simulations are compared with those of the PN. If we assume the exact parabolic orbit (EPO) without any changes, the radiated energy is given by \citep{quinlan87,quinlan89}
\begin{equation}\label{eq:quinlan_E}
\Delta E=\frac{85\pi}{12\sqrt{2}}\frac{G^{7/2}}{c^5}\frac{m_{1}^{2}m_{2}^{2}(m_{1}+m_{2})^{1/2}}{r_{\rm p}^{7/2}},
\end{equation}
where $G$ and $c$ are gravitational constant and speed of light, respectively, and $r_{\rm p}$ is the pericenter distance of the parabolic orbit. The GR capture will occur if the orbital energy is less than $\Delta E$. If we use the relation between $r_{\rm p}$ and angular momentum $L$ as $r_{\rm p}=(m_{1}+m_{2})L^{2}/2G m_{1}^2 m_{2}^2$, the Eq.~(\ref{eq:quinlan_E}) can be rewritten as
\begin{equation}\label{eq:delE_pn}
\Delta E = \frac{170\pi}{3} \frac{G^7}{c^5} \frac{m_{1}^{9} m_{2}^{9}}{(m_{1}+m_{2})^{3}} \frac{1}{L^7}~.
\end{equation}
Similarly, the radiated angular momentum $\Delta L$ can be expressed with the orbital angular momentum $L$,
\begin{equation}\label{eq:delL_pn}
\Delta L = 24\pi \frac{G^5}{c^5} \frac{m_{1}^{6}m_{2}^{6}}{(m_{1}+m_{2})^2} \frac{1}{L^4}~.
\end{equation}
Now we can compare these results directly with those from the numerical relativity (NR).

However, the exact parabolic orbit (EPO) is not possible in practice. During the encounter, the orbit changes gradually due to the gravitational radiation. In order to compare our numerical results with the PN approximation in a more consistent manner, we should take into account the corresponding changes of the orbit. 

\begin{figure}
\includegraphics[width=0.95\columnwidth]{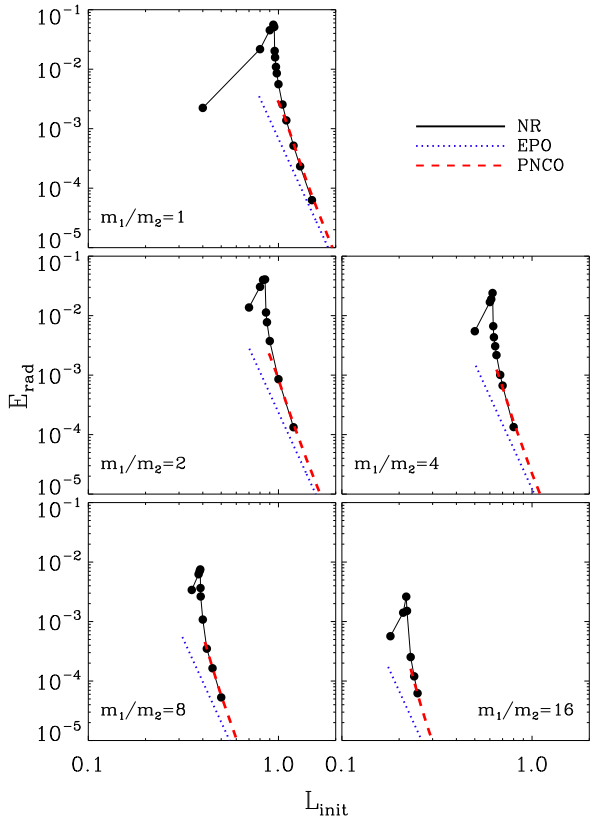}
\caption[Comparison with Post-Newtonian in radiated energy]{The comparison of the radiated energy from the relativistic simulations (solid lines) with PNs. Dotted lines are from the exact parabolic orbits (EPOs, Eq.~\ref{eq:delE_pn}) and the dashed lines are from the PN corrected orbits (PNCOs). \label{pn_radE}}
\end{figure}

\begin{figure}
\includegraphics[width=0.95\columnwidth]{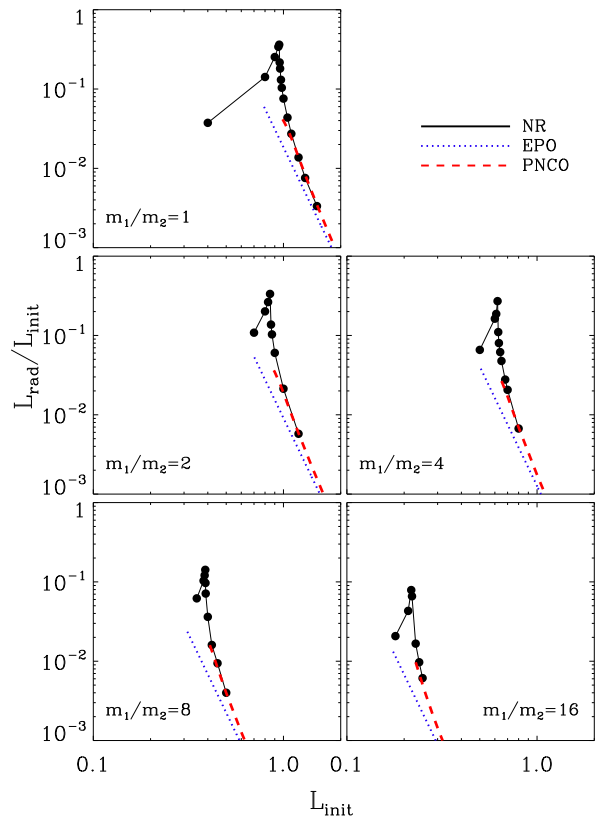}
\caption[Comparison with Post-Newtonian in radiated angular momentum]{The comparison of the radiated angular momentum from the relativistic simulations (solid lines) with PNs. Dotted lines are from EPOs (Eq.~\ref{eq:delL_pn}) and the dashed lines are from PNCOs. \label{pn_radL}}
\end{figure}

We have used a \textsc{toy} code\footnote{`ksreg2.tar.gz' in code group toy at: \\http://www.ast.cam.ac.uk/$\sim$sverre/web/pages/nbody.htm} for two body motions written by S. J. Aarseth and we have added the PN correction terms up to 3.5 PN order to compute the orbits more realistically for non-spinning case \citep{blanchet14}. The time derivatives of the acceleration terms should also be added because it adopts the Hermite integration schemes. Along the orbit, we have calculated the radiated energy and angular momentum by using the quadrupole formula \citep{peters64}. We adopted the initial separation of two BHs with $10^{4}$ M for the orbit calculations.

Figure~\ref{pn_orbit} shows the examples of the exact parabolic orbit (EPO), PN corrected orbit (PNCO) and NR orbit. Compared to the EPO, the PNCO and the NR orbits are more tightly wound. Near the pericenter where most GWs are generated, PNCO and NR orbits are nearly the same. Therefore, we can expect that the PNCO will provide more accurate values of the GW radiations than the EPO.

Figures ~\ref{pn_radE} and \ref{pn_radL} show the comparisons of the radiated energies and angular momenta between the NR and the PN. The radiated energy and angular momentum of EPO and PNCO are calculated up to the orbits with pericenter distance of 5 M.

Around the initial angular momentum of peak points, NR gives much larger radiated energy and angular momentum by an order of magnitude than EPO (dotted lines). On the other hand, the PNCOs (dashed lines) provide much more consistent results as expected from Fig.~\ref{pn_orbit}.

\subsection{Critical Impact Parameters for Capture \label{crtimpt}}

In the Newtonian limit, the angular momentum $L$ and the energy $E$ of the hyperbolic orbit are given as follows,
\begin{equation}\label{eq:hyper_l}
L=b\mu v_{\infty}~,
\end{equation}
\begin{equation}\label{eq:hyper_e}
E=\frac{1}{2}\mu v_{\infty}^{2}~,
\end{equation}
where $b$ is the impact parameter, $\mu$ is the reduced mass and $v_{\infty}$ is the relative velocity between two BHs at infinity. From these equations, the impact parameter can be expressed with the energy and angular momentum of the hyperbolic orbit as follows,
\begin{equation}\label{eq:impactp}
b=\frac{L}{\mu v_{\infty}}=\frac{L}{\sqrt{2\mu E}}~.
\end{equation}

\begin{figure}
\includegraphics[width=0.95\columnwidth]{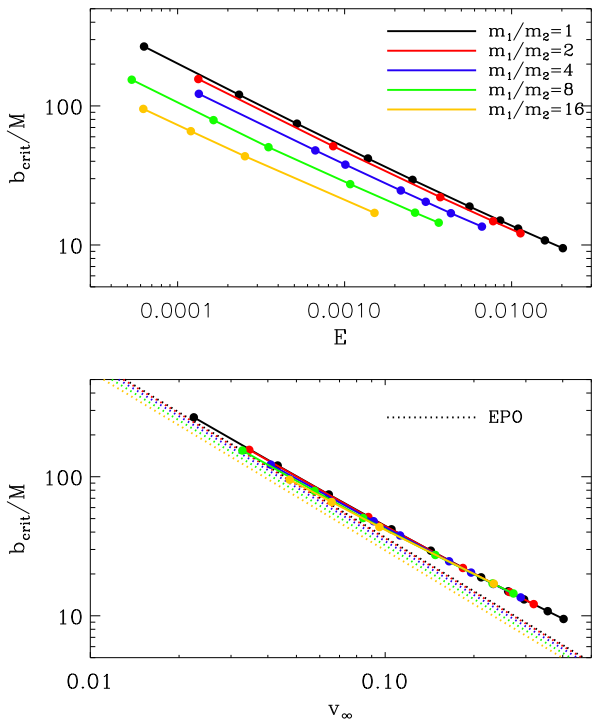}
\caption[Impact parameters from NR]{The critical impact parameters from NR simulations as a function of the energy $E$ (upper) and the relative velocity at infinity $v_{\infty}$ (lower). NR results deviate from the EPO's in high velocity region.\label{imp_ev}}
\end{figure}

Under the parabolic approximation, the critical impact parameter of the marginally capturing encounters can be calculated if we substitute the radiated energy $\Erad$ and initial angular momentum $\Linit$ in the parabolic orbit simulations for $E$ and $L$ in Eq. (\ref{eq:impactp}).

The upper panel of Fig.~\ref{imp_ev} shows the comparison of the critical impact parameters from NR simulations of different mass ratios as a function of $E$. Since it is more difficult to capture the BHs with higher energy, their critical impact parameters are smaller. As the mass ratio increases, the critical impact parameter decreases for a given energy.

It is more intuitive to express the critical impact parameters $\bcrit$ for capture in terms of $v_{\infty}$. For EPO, we get $\bcrit$ as a function of $v_{\infty}$ using Eq. (\ref{eq:delE_pn}), (\ref{eq:hyper_l}), and (\ref{eq:hyper_e}),
\begin{equation}\label{eq:impact_pn}
\bcrit = \bigg( \frac{340\pi}{3} \frac{m_{1}m_{2}(m_{1}+m_{2})^{5}}{v_{\infty}^{9}} \bigg)^{1/7}~.
\end{equation}
According to this equation, the critical impact parameter for capture has very weak dependence on the masses: $\bcrit \propto(m_{1}m_{2})^{1/7}$ if the total mass is fixed. NR results show even weaker dependence than this, i.e., nearly independent of the mass ratio (lower panel of Fig.~\ref{imp_ev}), and their slopes are less steep compared to the EPO's ($\bcrit \propto v^{-9/7}$). NR always gives larger impact parameters than EPO's especially in the high velocity region, and the deviation of NR from EPO becomes significant when $\bcrit \lesssim100$M.

\begin{figure}
\includegraphics[width=0.95\columnwidth]{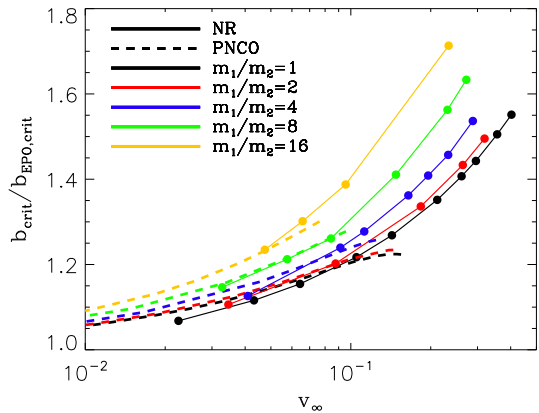}
\caption[The comparison of the impact parameters with different mass ratios from different methods]{The comparison of the critical impact parameters from different methods. The impact parameters are divided by those of the EPO. \label{impratio_v}}
\end{figure}

Figure~\ref{impratio_v} shows the critical impact parameters from PNCO and NR normalized by those of EPO. The EPO approximation which uses the results of \citet{peters64} gives consistent impact parameters with NR and PNCO up to $v_{\infty}\sim 0.01$ within about 10\%. Clearly EPO is suitable for the GR captures in globular clusters or nuclear star clusters \citep{quinlan89,oleary09,hong15} where the velocity dispersions are generally smaller than 0.01.

As the $v_{\infty}$ becomes larger, the critical impact parameters from the NR as well as the PNCO deviate more from EPO's. PNCO still gives reliable $\bcrit$ up to $v_{\infty}\sim0.1$ which corresponds to $\bcrit \sim40$ M in Fig.~\ref{imp_ev}, but it also deviates from NR beyond that range. The deviation is more conspicuous for the high mass ratios. For instance, the $v_{\infty}$ should be about 0.35 to reach 50\% of the deviation for the equal mass case while it can be achieved with $v_{\infty}\sim0.14$ for the mass ratio $m_{1}/m_{2}=16$. These value correspond to  $\bcrit \sim10$ M and $\bcrit \sim30$ M, respectively (Fig.~\ref{imp_ev}). 

The GR capture processes with higher velocity encounters require more accurate treatments for the relativistic effects. Their orbits deviate from EPO significantly. The general relativistic effects in this region should be considered more seriously, especially for the high mass ratio.


\section{\label{validity}Validity of Parabolic Approximation}

In this study, we have adopted the parabolic approximation which uses the parabolic orbit instead of the hyperbolic orbit for the simulations of GR capture. But obviously, the parabolic approximation implicitly assumes that the orbits are only weakly hyperbolic, i.e., orbital energy is much smaller than the rest-mass energy. Therefore, it is important to check the validity of the parabolic approximation for the high velocity encounters. As mentioned in section \ref{setup}, finding the hyperbolic orbit that gives marginal capture is very time consuming in NR because of the additional parameter (i.e., energy) to specify the orbits. Therefore, we will use the PN approaches which do not require heavy calculations for that.

For the exact hyperbolic orbit (EHO), similarly with the exact parabolic orbit (EPO), we can obtain the critical impact parameter of GR capture directly using \citet{hansen72}'s results by equating the initial orbital energy with the radiated energy (Eq. (13)--(17) in \citet{hansen72}).

The PN corrected hyperbolic orbit can be calculated in the same way with the parabolic orbit using the PN equation of motion. We have repeated the PNCO calculations until the difference between the initial orbital energy and the radiated energy becomes less than 0.01\% of the initial orbital energy.

\begin{figure}
\includegraphics[width=0.95\columnwidth]{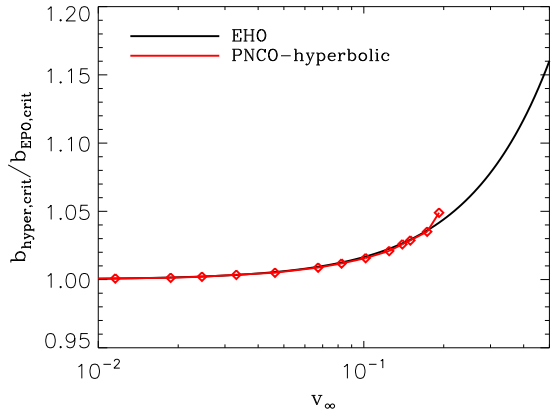}
\caption[The impact parameters of hyperbolic orbit with respect to the parabolic orbit]{The impact parameters from the exact hyperbolic orbit (EHO) and PN corrected hyperbolic orbit (PNCO-hyperbolic) with respect to the exact parabolic orbit (EPO). These are equal mass and non-spinning cases. \label{eq_imp_hyper}}
\end{figure}

Figure~\ref{eq_imp_hyper} shows the critical impact parameters from the exact hyperbolic orbit (EHO) and PN corrected hyperbolic orbit (PNCO-hyperbolic) normalized by that of exact parabolic orbit (EPO). We found that $b_{\rm hyper,crit}/b_{\rm EPO,crit}$ remains below 1.05 for both EHO and PNCO-hyperbolic cases up to $v_{\infty}\sim0.2$. This means the parabolic approximation is valid within 5\% for that $v_{\infty}$ range. For the encounters with even higher velocity, the hyperbolic orbit gives larger deviations from the parabolic orbit, which implies that true critical impact parameters of hyperbolic orbit can be at least 5\% larger than that based on the parabolic approximations at $v_{\infty}\gtrsim0.2$. Even though we cannot examine the NR for the true hyperbolic orbits, NR is expected to give qualitatively similar results because PNCO has shown good agreements with NR for the parabolic orbits.


\section{\label{discuss}Discussions}

We carried out detailed studies of GR capture between two BHs using NR. We employed parabolic approximation for the initial orbits and found that such an approximation is applicable for the encounters with relative velocity up to $10\sim20$ \% of the speed of light. 

The velocity dispersion can be very large in the vicinity of the supermassive BHs. \citet{oleary09} argued that the encounters between relatively massive stellar mass BHs near the supermassive BH of the galactic nuclei can be quite frequent. Some of them would occur with very large relative velocity. Therefore, the galactic center near the supermassive BH is the most probable place of the  relativistic encounters.


The signature of the strong relativistic encounters would be the GWs coming from very eccentric mergers. Such events would be rare but even a few events will tell us about the physical conditions where they were born.

The GR capture with very high mass ratio such as the supermassive BH and the stellar mass BH also requires more accurate relativistic treatments. The critical impact parameter between them can be much larger than what we expect in EPO. The mass ratios in this study are limited to 16, but we have seen that the deviation of critical impact parameter of high mass ratio from EPO begins at lower relative velocity, and is getting larger than that of low mass ratio. Therefore, the EPO should be used very carefully for those cases even though their relative velocity is not larger than 0.01. 

In the present study we have not considered spins of the BHs. In very close encounters between BHs, spin is expected to play important roles. We will return to this subject in the forthcoming paper.

\section{Acknowledgments}

HML was supported by NRF grant number NRF-2006-0093852 funded by the Korean government and partially by the KISTI-GSDC's Cooporative Program (K-16-L01-C06-S01). GK was supported in part by the R\&D Program of KISTI (K-16-L05-C01-S01,K-17-L01-C02-S04) and the Academic Program of APCTP. The computation was carried out with the supercomputer and technical supports in KISTI through the HPC application program (KSC-2014-C3-017, KSC-2014-C3-065). We also thank Heeil Kim and Jongsuk Hong for the advice about the code and manuscript.

\bibliography{GRcap_uneq.bib}

\end{document}